\documentclass[sigconf,anonymous=false]{acmart}
\usepackage[colorinlistoftodos]{todonotes}
\usepackage{xspace}

\usepackage{multirow}
\usepackage{multicol}
\usepackage{makecell}
\usepackage{subfig}
\usepackage{slashbox}
\usepackage[ruled,linesnumbered]{algorithm2e}

\usepackage{acronym}
\usepackage{arydshln}
\usepackage{booktabs}

\usepackage{mathrsfs}
\usepackage{mathtools}
\usepackage[inline]{enumitem}
\usepackage{xcolor, soul}
\usepackage{tikz}

\usepackage{balance}

\usetikzlibrary{shapes.misc, positioning}

\sethlcolor{lightgray}

\newcommand{\xamazon}{XMarket\xspace}
\newcommand{\ourmodel}{FOREC\xspace}
\newcommand{\partitle}[1]{\vspace{1mm}\noindent\textbf{#1}}
\newcommand{\market}[1]{\textbf{\texttt{#1}}}
\newcommand{\category}[1]{\texttt{#1}}

\acrodef{CF}{Collaborative Filtering}
\acrodef{CDR}{Cross-Domain Recommendation}
\acrodef{CMR}{Cross-Market Recommendation}
\acrodef{MAML}{Model-Agnostic Meta-Learning}
\acrodef{GMF}{Generalized Matrix Factorization}
\acrodef{MLP}{Multi-Layer Perceptron}
\acrodef{NMF}{Neural Matrix Factorization}
\acrodef{HR}{Hit-Rate}
\AtBeginDocument{%
  \providecommand\BibTeX{{%
    \normalfont B\kern-0.5em{\scshape i\kern-0.25em b}\kern-0.8em\TeX}}}

\setcopyright{acmcopyright}

\copyrightyear{2021}
\acmYear{2021}
\setcopyright{acmcopyright}\acmConference[CIKM '21]{Proceedings of the 30th ACM International Conference on Information and Knowledge Management}{November 1--5, 2021}{Virtual Event, QLD, Australia}
\acmBooktitle{Proceedings of the 30th ACM International Conference on Information and Knowledge Management (CIKM '21), November 1--5, 2021, Virtual Event, QLD, Australia}
\acmPrice{15.00}
\acmDOI{10.1145/3459637.3482493}
\acmISBN{978-1-4503-8446-9/21/11}

\begin{document}
\fancyhead{}

\title[Cross-Market Product Recommendation]{Cross-Market Product Recommendation}

\author{Hamed Bonab}
\affiliation{%
  \institution{University of Massachusetts Amherst} 
}
\email{bonab@cs.umass.edu}

\author{Mohammad Aliannejadi}
\affiliation{%
  \institution{University of Amsterdam} 
}
\email{m.aliannejadi@uva.nl}

\author{Ali Vardasbi}
\affiliation{%
  \institution{University of Amsterdam}
}
\email{a.vardasbi@uva.nl}

\author{Evangelos Kanoulas}
\affiliation{%
  \institution{University of Amsterdam}
}
\email{e.kanoulas@uva.nl}

\author{James Allan}
\affiliation{%
  \institution{University of Massachusetts Amherst}
}
\email{allan@cs.umass.edu}

\renewcommand{\shortauthors}{Bonab, et al.}

\begin{abstract}
We study the problem of recommending relevant products to users in relatively resource-scarce markets by leveraging data from similar, richer in resource auxiliary markets. We hypothesize that data from one market can be used to improve performance in another.
Only a few studies have been conducted in this area, partly due to the lack of publicly available experimental data. 
To this end, we collect and release XMarket, a large dataset covering $18$ local markets on $16$ different product categories, featuring $52.5$ million user-item interactions. 

We introduce and formalize the problem of cross-market product recommendation, i.e., \textit{market adaptation}. 
We explore different market-adaptation techniques inspired by state-of-the-art domain-adaptation and meta-learning approaches and propose a novel neural approach for market adaptation, named \textit{FOREC}. Our model follows a three-step procedure -- pre-training, forking, and fine-tuning -- in order to fully utilize the data from an auxiliary market as well as the target market. We conduct extensive experiments studying the impact of market adaptation on different pairs of markets. Our proposed approach demonstrates robust effectiveness, consistently improving the performance on target markets compared to competitive baselines selected for our analysis. 
In particular, FOREC improves on average 24\% and up to 50\% in terms of nDCG@10, compared to the NMF baseline.    
Our analysis and experiments suggest specific future directions in this research area. We release our data  and code\footnote{Data and code: \url{https://xmrec.github.io}} for academic purposes. 
\end{abstract}



\maketitle

\section{Introduction}
Nowadays online shopping in many countries is a part of people's daily lives. While online shopping brings several benefits and comfort to both users and vendors~\cite{jiang2013measuring}, it comes at the risk of overwhelming users with virtually unlimited options to choose from. Recommender systems are key in dealing with information overload, helping not only users finding interesting items, but also vendors finding the right customer for their products. E-commerce companies often operate across markets; for instance Amazon\footnote{\url{https://www.amazon.com}} has expanded their operations and sales to 18 markets around the globe.\footnote{\url{https://sell.amazon.com/global-selling.html}} This brings both opportunities and challenges. 

While it is typical that several local e-commerce companies operate in every country, the presence of an international e-commerce company, like Amazon, eBay, and Etsy can benefit users even more if these companies can utilize the experience and data gathered across several markets.
Using cross-market data however comes at a risk of assuming one-solution-fits-all and applying the same algorithms that are developed for and trained on large and data-rich markets, such as the U.S.~\cite{im2007does}, to small and data-scarce markets. The key challenge is that data, such as user interaction data with products (clicks, purchases, reviews), convey certain biases of the individual markets~\cite{DBLP:conf/sigir/CanamaresC18}. 
Algorithms that are optimized for a certain market learn various biases and distributions of the data~\cite{DBLP:conf/www/SunKNS19}.
Therefore, the algorithm trained on a source market, are not necessarily effective in a different target market~\cite{ferwerda2016exploring}, since utilizing the vast amount of data from a large market to improve the performance on low-resource markets comes at the risk of importing the wrong data distributions.
For example, assume iPhone is the most popular smartphone in the U.S. (source market), while Samsung is the most popular smartphone brand in Germany (target market). Importing user preference from the U.S.~market would yield to recommending iPhone in Germany more often than Samsung, which clearly is a wrong choice.
Hence, even though there is a myriad of information to learn from a source market, careful adaptation of data is required. \looseness=-1

\clubpenalty=0
\widowpenalty=0

The significance of \ac{CMR} has been pointed out in the literature~\cite{roitero2020leveraging}. However, small progress has been made in this area, mainly due to a lack of experimental data.
To this end, we construct a large-scale real-life product recommendation dataset, referred to as \xamazon, drawn from $18$ markets in $11$ languages. 
To develop this dataset we crawled Amazon marketplaces around the globe, locating and including in the dataset the same products within different markets. 
Moreover, we analyze certain statistical properties and trends amongst multiple markets and product categories where we highlight the existence of crucial differences across different markets. 

In this paper, we focus on the user-item interaction data through ratings and study the problem of recommending relevant products to users in relatively resource-scarce markets by leveraging data from similar, richer in resource auxiliary markets. Our hypothesis is that data from one market can be used to improve performance in another.
For this purpose, we first introduce and formalize the problem of cross-market product recommendation, i.e., \textit{market adaptation}. 
In order to solve the problem of \ac{CMR}, we explore market-adaptation baselines inspired by domain-adaptation and meta-learning approaches. Then, we propose a novel neural approach, named \textit{FOREC}, consisted of a three-step procedure -- pre-training, forking, and fine-tuning -- in order to fully utilize the data from an auxiliary market to boost the product recommendation performance in the target market. 
More specifically, FOREC learns a general recommendation system based on two markets (i.e., source and target) and employs a forking procedure by adding a \textit{market specific} sub-network to the head of the model and freezing the bottom part in order to adapt the general internal representations to the target market. 
We conduct extensive experiments studying the impact of market adaptation on different pairs of markets. FOREC demonstrates robust effectiveness, consistently improving the performance compared to competitive baselines on $7$ target markets we selected for our study. 
In particular, FOREC improves on average 24\% and up to 50\% in terms of nDCG@10, compared to the NMF baseline. Our analysis and experiments provide many insights on the \ac{CMR} and suggest specific future directions in this research area.    
In summary, the main contributions of this paper constitute:
\begin{itemize}[leftmargin=*]
    \item Collecting a real-world pragmatic large-scale cross-market and cross-lingual product review dataset.
    \item Performing analysis of cross-market behavioral biases pivoting on the fact that our dataset includes the same items shared across different markets. In particular, we study how differently users in different markets interact with the same products.
    \item Proposing a novel neural architecture for market-adaptation and  demonstrating the effectiveness of the model through extensive experiments. We adapt various \acf{CDR} and meta-learning approaches for \acf{CMR}.
    \item Analyzing the performance of the model, considering various setups and conditions to provide further insights.
\end{itemize}

\section{Related Work}
\label{sec:relatedwork}
This study is related to \ac{CMR} and \ac{CDR}, as well as meta-learning approaches. In this section, we briefly review the research done in these domains. 

\partitle{Cross-domain and cross-market recommendation.}
Research on \ac{CMR} and \ac{CDR} aims at improving the system's effectiveness based on the external data that is available from other markets or item categories. 
While having the same goal, the two tasks differ in various aspects, each bearing their own challenges. 
Particularly, in \ac{CDR} the general assumption is that the model learns from interactions of overlapping users
\footnote{with a few exceptions like~\citep{perera2019cngan, DBLP:conf/sigir/KrishnanDBYS20} that assume non-overlapping users} 
in different domains 
(e.g. categories in product search) 
with the aim of improving the recommendation on the target domain items, using help from the source domain items.
For \ac{CMR} the situation is reversed: interactions of different set of users in the source market are leveraged to boost the recommendation for users in the target market. Here, we assume that the items are shared among different markets.

\citet{im2007does} conduct an experiment in two product domains, aiming to answer the question ``does a one-size recommendation fit all?'' where they observed that the performance of \ac{CF} is highly affected by the information-seeking mode of the users. 
Depending on the product domain, users adopt different strategies and therefore the system would not fit to all domains. 
\citet{lu2013selective} later argue that transferring all the knowledge from source domain into the target domain may harm the recommender due to some inconsistencies and propose a criterion for selecting the consistent part of knowledge to be transferred to the target domain. 
\citet{DBLP:conf/www/ElkahkySH15} apply domain adaptation using user behavior-based features for learning latent space.
\citet{DBLP:conf/cikm/RafailidisC17} propose a collaborative ranking model with a weighting strategy that controls the influence of user preferences from auxiliary domains. 
\citet{zhao2020catn} use reviews to transfer user preference at aspect-level as a cross-domain recommendation framework. 
\citet{DBLP:conf/sigir/KrishnanDBYS20} propose using contextual invariances across domains to leverage data from a dense domain to improve learned representations in other sparse domains.
Different from these works, in this paper we investigate the existence of different behavioral biases across different markets where unlike \ac{CDR}, the items are the same across different markets but the users are different. 
Further research has aimed at mitigating these biases and transferring knowledge from one domain to another~\cite{hu2018conet, li2020ddtcdr}. 
Unsupervised domain adaptation \citep{ganin2015unsupervised} has also inspired various cross-domain recommender systems in recent years~\citep{kanagawa2019cross, wang2019recsys, li2020atlrec, yuan2019darec}.
\ac{CDR} has been used specifically for mitigating the cold-start problem in a number of studies ~\citep{jiang2015social, mirbakhsh2015improving, wang2018cross, kang2019semi, fu2019deeply, zhao2020catn, jin2020racrec}.

While there exists much work on domain adaptation, \emph{market-adaptation} is relatively unstudied. \ac{CMR} has attracted attention in music recommendation~\cite{ferwerda2016exploring,roitero2020leveraging} where \citet{ferwerda2016exploring} analyze and study music diversity across countries and propose to use country-based diversity measurements for system evaluation. \citet{roitero2020leveraging} studied user behavior in 21 different markets on Spotify and highlight the need for market-specific algorithms, as opposed to a global algorithm. We take one step further in this direction by expanding our study to various item categories in e-commerce where users purchase items (rather than having a monthly subscription) and express their opinion and experience with item in the form of ratings and reviews.

\partitle{Meta-learning.}
The goal of meta-learning is to train a model on multiple tasks, such that it can rapidly adapt to a new task after seeing a small number of new training samples~\citep{vilalta2002perspective}.
In the context of recommendation systems, meta-learning has been used for several problems, including but not limitted to recommender algorithm selection~\citep{collins2018novel, cunha2018metalearning, luo2020metaselector}, cold-start problem~\citep{vartak2017meta, lee2019melu}, and retraining the model~\citep{zhang2020retrain}.
For meta-learning of deep neural networks, a general and powerful technique, \ac{MAML}, has been proposed that can be directly applied to any learning problem and model~\citep{finn2017model}.
\ac{MAML} framework is widely used in recommendation literature.
For example, \citet{lu2020meta} use it on heterogeneous information networks to address the cold-start problem.
Others used \ac{MAML} to train a recommender which performs reasonably good enough both for cold- and warm-start users \citep{bharadhwaj2019meta, lee2019melu}.

\setlength{\textfloatsep}{0.2cm}
\begin{table}[]
   \vspace{-3mm}
    \centering
    \caption{General statistics of \xamazon.}
    \vspace{-1em}
    \begin{tabular}{ll}
    \toprule
        \# markets &  18 \\
        \# languages & 11 \\
        \# categories & 16 \\
        \# items & 3,811,438 \\
        \# users & 9,562,260\\
        \# reviews & 52,480,184\\
        \# reviewed items & 1,000,829 \\
        \# unique items & 294,739 \\
    \midrule
    \end{tabular}
    \label{tab:stats}
\end{table}

\section{Data Collection \& Analysis}

In this section, we first describe in detail the data collection process and provide statistics for the collected data. Then, we analyze the data and highlight important characteristics and similarities across different markets.

\subsection{Data Collection}
We describe our new dataset, called \xamazon, and provide details on how we generated it. We constructed the \xamazon item and review collection on top of a large-scale publicly-available Amazon dataset~\cite{DBLP:conf/sigir/McAuleyTSH15,DBLP:conf/www/HeM16}\footnote{\url{https://jmcauley.ucsd.edu/data/amazon/}}. 
The Amazon Product data~\cite{DBLP:conf/sigir/McAuleyTSH15} includes millions of item reviews collected from the Amazon U.S.~marketplace in various categories. The dataset was collected in 2014 and later updated in 2018~\cite{DBLP:conf/emnlp/NiLM19}. We used this dataset as a seed to initiate our crawl. We located the same items that appear on the U.S.~market in other markets, by matching the items' unique identifiers (aka.~ASIN's) on all available Amazon markets. 
In particular we have crawled data from the following markets: Saudi Arabia (\market{sa}), Singapore (\market{sg}), Australia (\market{au}), United Arab Emirates (\market{ae}), Turkey (\market{tr}), Japan (\market{jp}), India (\market{in}), Spain (\market{es}), U.S.~(\market{us}), China (\market{cn}), Germany (\market{de}), Netherlands (\market{nl}), France (\market{fr}), Brazil (\market{br}), Canada (\market{ca}), Mexico (\market{mx}), Italy (\market{it}), United Kingdom (\market{uk}).
Our main criterion for including an item in the collection process was its popularity on the \market{us}~market. 
Our decision was motivated by the idea of having a high-resource market (i.e., \market{us}) that would provide a wealth of data to other markets. Therefore, we discarded all items with less than 20 reviews in the past two years, as we did not consider them rich enough to be useful in other markets. 
In our preliminary analysis, we noticed that in most cases, if an item exists in another market its ASIN is the same. Therefore, we fed our crawler with the ASIN's that we collected from the U.S.~market. In doing so, we collected cross-market metadata information for over one million items and collected about 52 million multilingual reviews.

Among the existing multi-lingual review datasets, we find the Multilingual Amazon Reviews Corpus (MARC) \cite{marc_reviews} the most similar to \xamazon. MARC consists of multi-lingual reviews extracted from different Amazon marketplaces, however, the scale of the dataset is much smaller. In particular, they do not cover all Amazon marketplaces and categories, whereas \xamazon covers a wide range of categories in all 18 Amazon marketplaces. Moreover, \xamazon includes rich item and review metadata (e.g., reviewer ID, item description, and related items) that can be utilized to pursue various research directions. 
We also found another similar dataset named as Amazon Customer Reviews Dataset\footnote{\url{https://s3.amazonaws.com/amazon-reviews-pds/readme.html}} providing a collection of reviews from five marketplaces dated from 1995 to 2015. We notice that a vast majority of the provided data is only from United States and the provided meta-data is limited to only product title and reviews whereas our data covers more number of marketplaces with a full meta-data information (including product text, images, also bought, similar items). In addition, our product reviews are more recent.
To the best of our knowledge, no other cross-market multi-lingual recommendation dataset with such a wide coverage of markets and categories exists in the community.

\subsection{Data Statistics}
\begin{figure*}
    \centering
    \subfloat[Distribution of items per market.]{\includegraphics[trim=0 0 0 0, clip,height=3.2cm,]{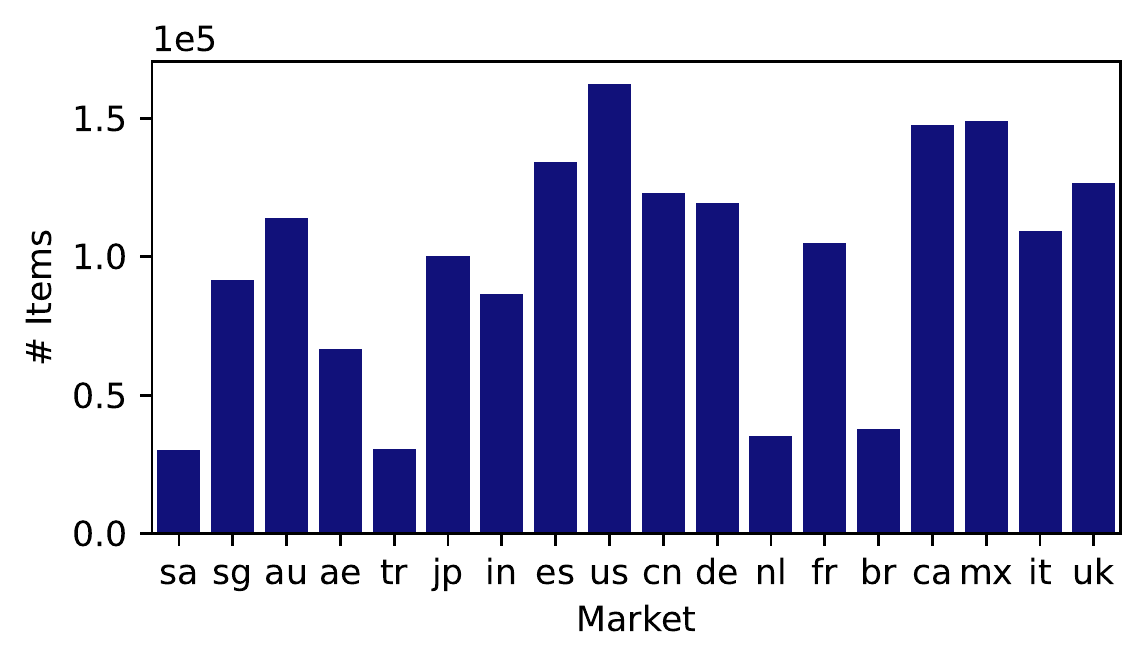}\label{fig:item_per_market}}
    \subfloat[Purchase Count]{\includegraphics[height=3cm,trim=0 0 60 0, clip]{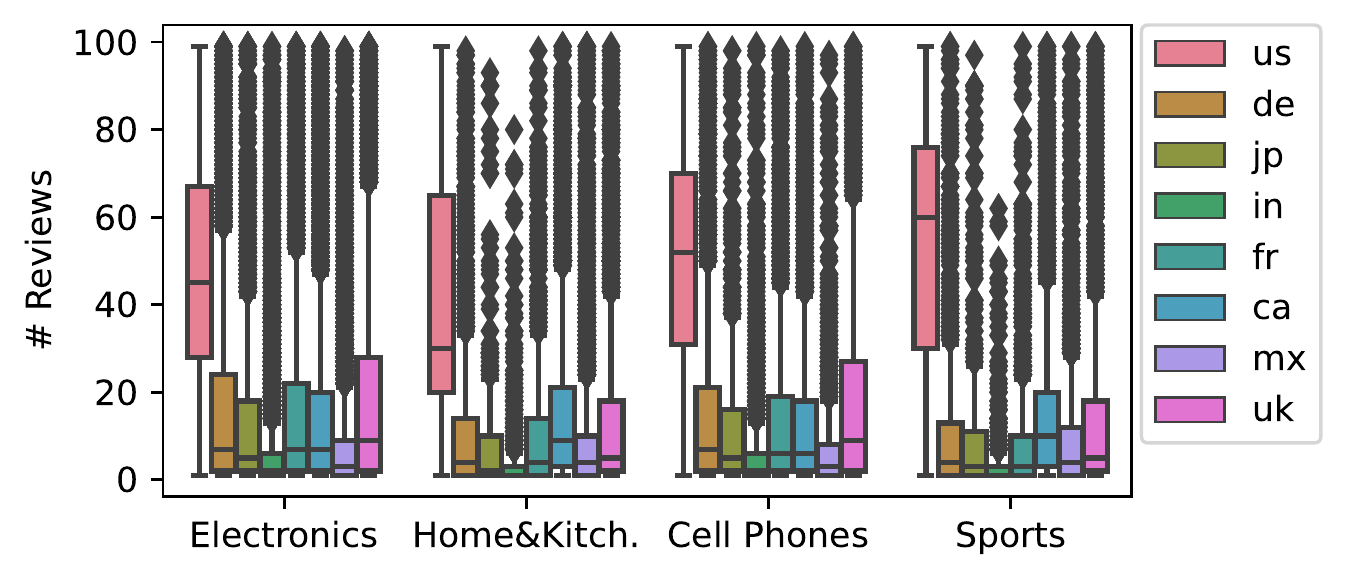}\label{fig:count_dist}}
    \subfloat[Rating Stars]{\includegraphics[height=3cm, trim=0 0 0 0, clip]{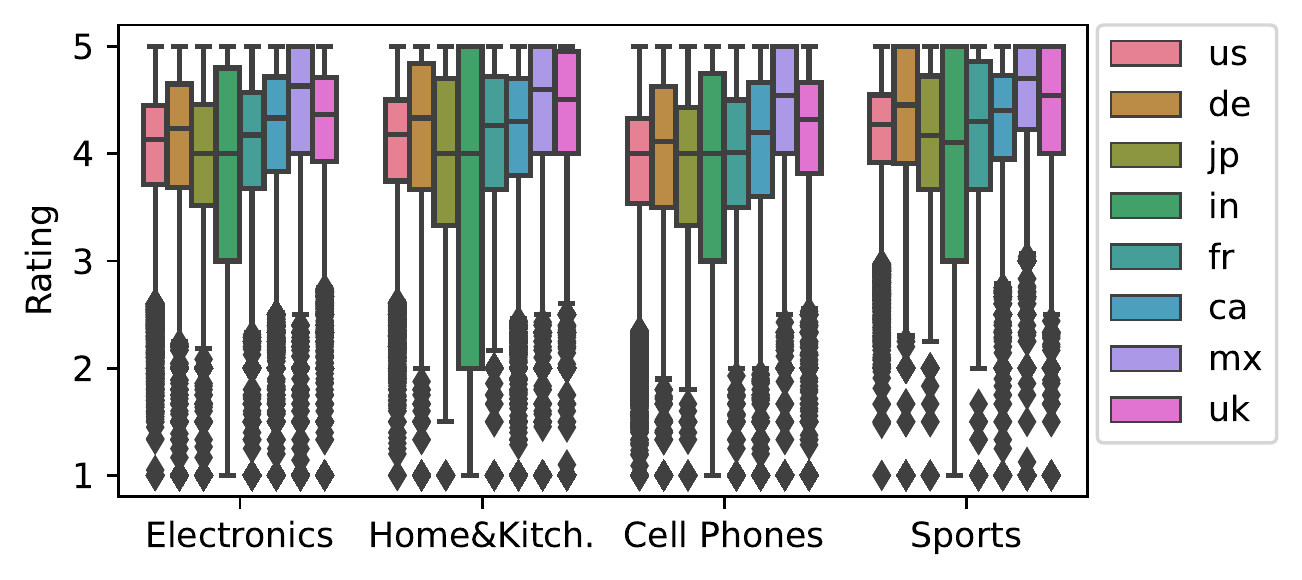}\label{fig:star_dist}}
    
    \vspace{-1em}
    \caption{Distribution of items, purchase count, and rating stars, per category and per market.}
    \label{fig:3_dists}
\end{figure*}

Table~\ref{tab:stats} summarizes some of the main characteristics of \xamazon. It provides a cross-lingual e-commerce dataset of 16 shopping categories in 11 languages. We crawled data for $\sim$300K unique items across all markets, which resulted in $\sim$4M cross-market items. 
Also, Fig.~\ref{fig:item_per_market} shows the distribution of items in each market. We see that the Canadian (\market{ca}) and Mexican (\market{mx}) markets have the most items in common with the U.S.~market (\market{us}), which is expected due to the long-lasting presence of Amazon in these countries and their vicinity to \market{us}.
Due to space considerations and similarity in results, in the remainder of the paper we analyze and discuss a subset of markets and categories. Our experiments and model evaluations are based on the \category{Electronics} category with statistics presented in Table~\ref{tab:marketstats}. We see in the table the main characteristics of the studied marketplaces in terms of recommendation data, such as number of items, users, and ratings. We observe a relatively high number of users as well as items in the U.S.~market, making it the most sparse market in our dataset. \looseness=-1

\setlength{\textfloatsep}{0.2cm}
\begin{table}[]
    \centering
    \vspace{-3mm}
    \caption{Statistics of different markets for \category{Electronics}.}
    \label{tab:marketstats}
    \vspace{-1em}
    \resizebox{1.0\columnwidth}{!}{
    \begin{tabular}{lllllllll}
    \toprule
        & \market{us} & \market{de} & \market{jp} & \market{in} & \market{fr} & \market{ca} & \market{mx} & \market{uk} \\
    \midrule
        \textbf{U} & 2.7m & 345k & 117k & 208k & 230k & 422k & 100k & 545k \\
        \textbf{I} & 35k &	8k &	4k &	7k &	6k & 	19k &	9k	& 10k \\
        \textbf{R} & 4.1m &	0.5m &	174k & 	257k & 	392k &	721k &	169k &	865k \\
         \bottomrule
    \end{tabular}
    }
\end{table}

\subsection{Cross-Market Analysis}
Our goal in this section is to analyze and demonstrate the similarities and differences across markets. Having in mind that the markets share the same set of products, we analyze how people in different regions interact with these items to uncover behavioral characteristics and biases.

\partitle{Distribution of ratings.}
Several reasons may influence users to purchase a product such as users' financial status, culture, and companies' marketing strategies.
Therefore, we study the difference in product rating (as a signal of product purchase) among different markets. We plot the distribution of product ``purchase'' in Fig.~\ref{fig:count_dist}. The dominance of \market{us} is obvious in this figure, with the highest median in all categories. We see that different markets exhibit different distributions across categories. In general we see that Electronics is most popular category among different markets. We also observe that the \category{Home \& Kitchen} category shows a different trend compared to other categories, perhaps because such items are more regionally dependent.

\partitle{Distribution of rating starts.}
We are interested in finding out if the same items are rated differently in each market. Also, if the differences happen across categories. We find significantly different behavior in giving rating stars among markets and categories, as shown in Figure~\ref{fig:star_dist}. We see a greater tendency of giving higher rating to items in \market{mx} market, whereas for \market{in} market, we see an opposite behavior. Interestingly, we observe a relatively similar trend in all categories where for example the median rating in \market{de} market is always higher than that of \market{us} and \market{fr} market, but slightly lower than the \market{uk} market. This clearly shows a general bias in user rating behavior, which should be taken into account when developing algorithms such as rating prediction. 

\begin{figure}[t]
    \centering
    \vspace{-5mm}
    \subfloat[Electronics]{\includegraphics[width=.5\columnwidth, trim=0 0 35 35, clip]{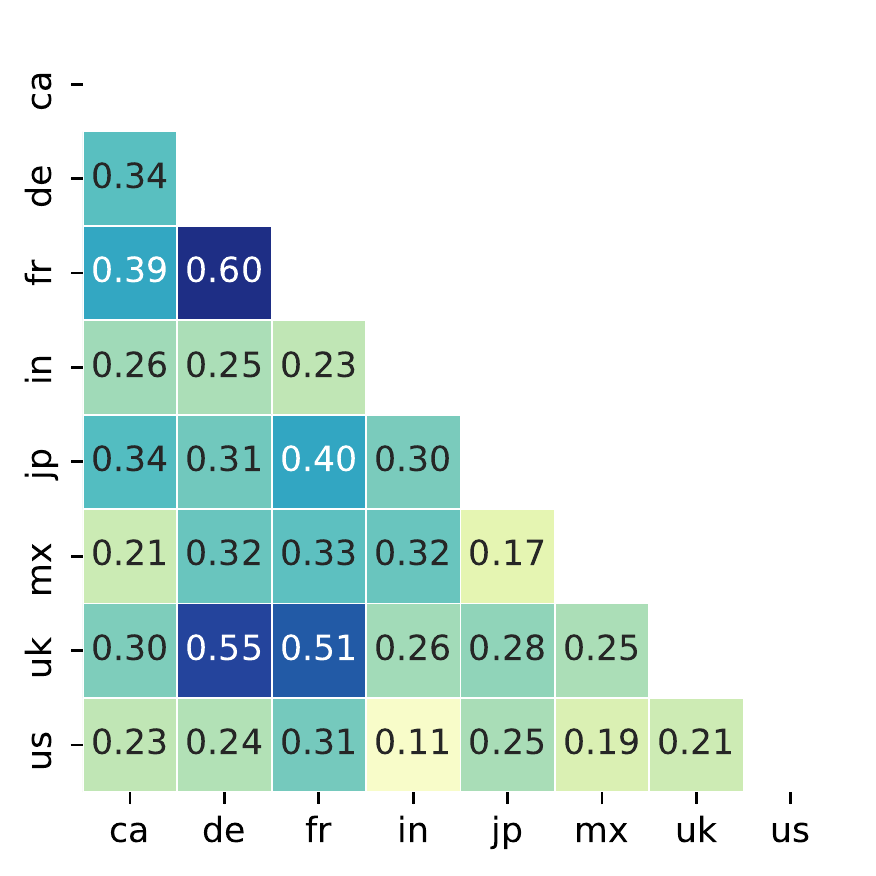}}
    \subfloat[Home \& Kitchen]{\includegraphics[width=.5\columnwidth, trim=0 0 35 35, clip]{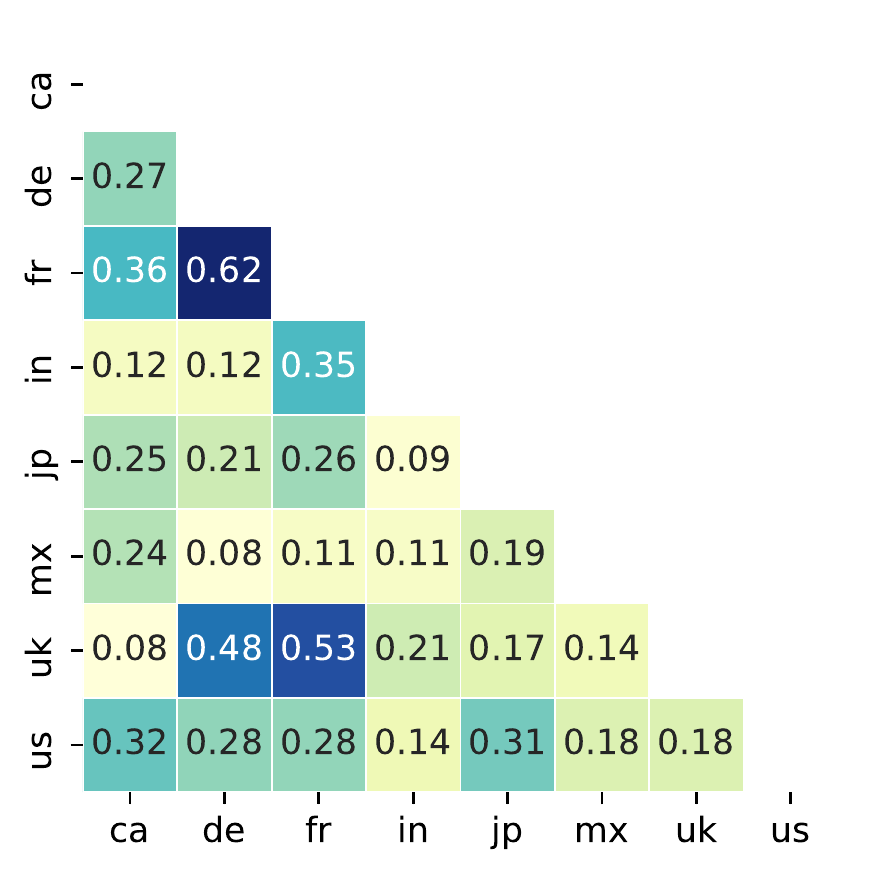}}
    \vspace{-2mm}
    \caption{Market similarity based on the cosine distance.}
    \label{fig:cosine_sim}
\end{figure}

\partitle{Market similarity.}
Based on the observations that we had from Figure~\ref{fig:3_dists}, we estimate the users' purchase similarity between markets. For a pair of markets in the same category, we build item purchase count vectors and compute cosine similarity between the two vectors. In Figure~\ref{fig:cosine_sim} we plot the similarities of all market pairs for two
categories. Interestingly, we observe the highest rate of similarity between \market{de}, \market{fr}, and \market{uk}, highlighting the similarities in European markets. On the other hand, we see that the American countries do not share much in common. As we see \market{us}, \market{ca}, \market{mx} exhibit low similarity, which is surprising. Perhaps this is due to strong existence of local vendors in this category. In particular, we see the lowest similarity between \market{de} and \market{mx}.

\partitle{Remarks.}
Overall, from our analyses it is evident that users in each market exhibit different behavior. These differences could be due to various reasons such as cultural biases or different marketing strategies adopted by companies. Another reason could simply be the popularity of Amazon as an e-commerce marketplace in different countries and how long it has been doing business in each country. For instance, we saw a very high similarity between \market{us}, \market{ca}, and \market{mx} in terms of common items that exist in the three markets, however, when it comes to product purchase data, we see very little similarity. This indicates that even though Amazon has a big influence on the e-commerce market in these countries, users act differently. 
With the existence of obvious differences and similarities, but at the same time having a mix of data-rich and scarce markets, learning from auxiliary markets is not trivial and requires careful development of market adaptation techniques.

\section{Problem Statement}

Assume we are given a set of parallel markets as $M = \{\mathbb{M}_0,\cdots,\mathbb{M}_t\}$. Let $\mathbb{M}_0$ be the base market with the set of items $\mathcal{I}_0=\{I_1,\cdots,I_n\}$. For the base market, one could assume the market with long-lasting existence offering the super-set of items and rich user-item interaction data. 
For example, with our XMarket settings, the \market{us} market can be thought of the base market and others such as \market{de} or \market{in} are considered the parallel target markets. We assume that $\mathcal{I}_l \subset \mathcal{I}_0$ for $1 \leq l \leq t$. Depending on the provided parallel markets, $M$, the base market could change or there might be no base market. With any of these settings, a union set of items in all the parallel markets could be defined as $\mathcal{I}_0$, satisfying our assumption.

For a given target market, $\mathbb{M}_l$, let the set of market's users as $\mathcal{U}_l=\{U^1_l,\cdots,U^z_l\}$. Generally, a user can interact with different markets, but for simplicity, we assume that the set of users in each market are mutually disjoint with any other parallel market. The problem of market adaption is to use any of the parallel markets provided, $\mathbb{M}_a \in M-\{\mathbb{M}_l\}$ as an auxiliary market to improve the quality of items recommended to users of the target market, i.e. $\mathcal{U}_l$. It is straightforward to use more than single auxiliary market. However, we focus on single auxiliary market and leave the other variations as the future work. For our experiments, we either augment with only \market{us} market or any of the parallel markets and report the results. Automatically selecting the most suitable parallel auxiliary market is another interesting problem that is out of the scope of this study and we plan to explore on that direction as our future work.

\setlength{\textfloatsep}{0.4cm}
\begin{figure}[t]
    \centering
    \includegraphics[width=\linewidth]{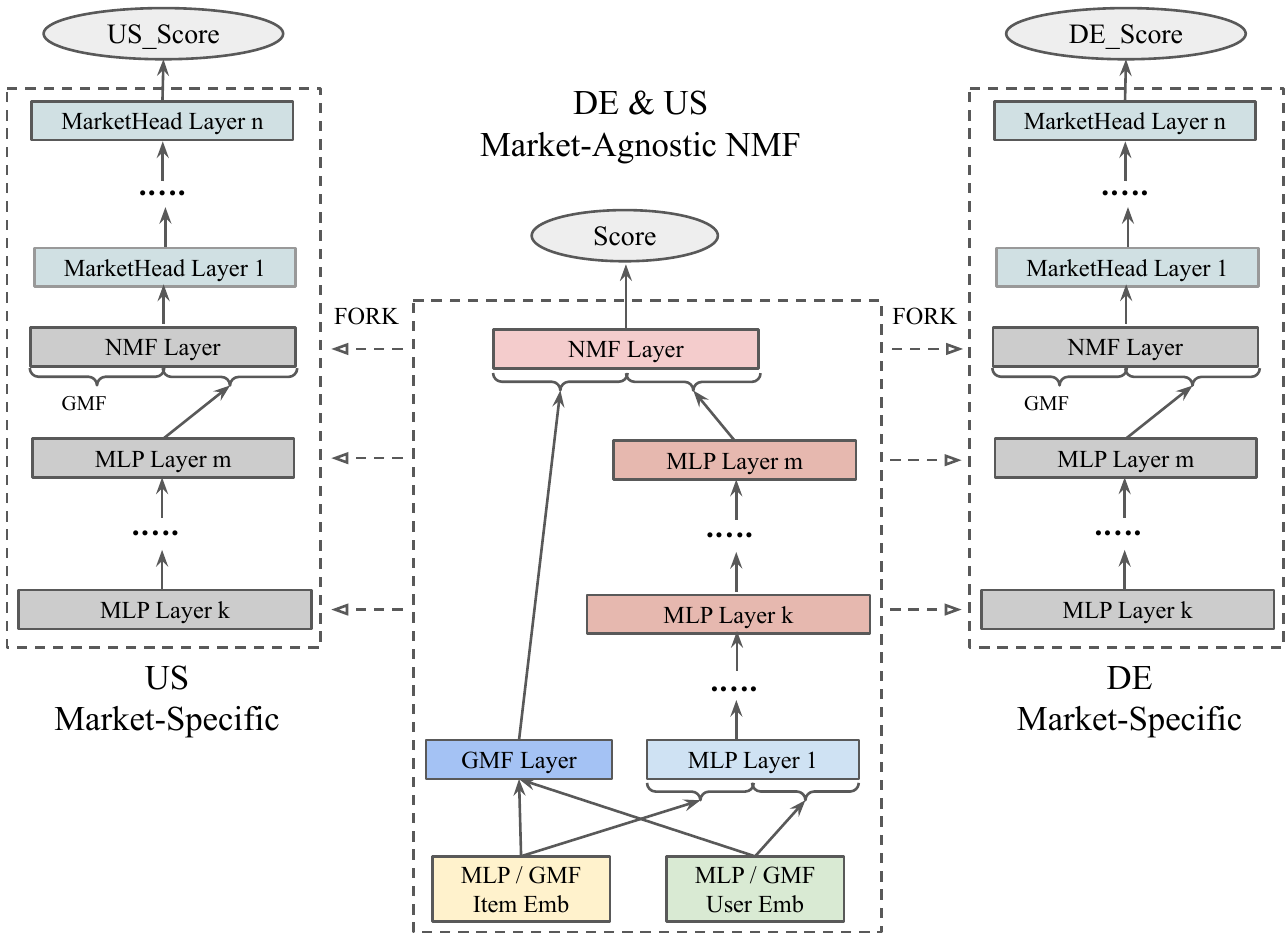}
    \vspace{-2em}
    \caption{The general schema of our FOREC recommendation system. For a pair of markets, the middle part shows the market-agnostic model that we pre-train, and then fork and fine-tune for each market shown in the left and right.}
    \vspace{-0.5em}
    \label{fig:forec}
\end{figure}

\vspace{-0.3em}
\section{FOREC: A CMR System}
\label{sec:model}

Here, we explain our proposed \ac{CMR} system, named \textit{FOREC}. A general schema of our model is presented in Fig. \ref{fig:forec}. We show an example model architecture for a pair of markets, \market{de} and \market{us}. The training phase for the FOREC system includes three ordered steps (i.e., pre-training, forking, and fine-tuning) that we explain in the following. Note that FOREC is capable of working with any desired number of target markets. However, for simplicity, we only experiment with pairs of markets for our experimental evaluation. \looseness=-1 

\subsection{Pre-Training: Market-Agnostic NMF}
In this step, we aim to train a recommendation model that is market-agnostic, in the sense that all the model parameters $\theta$ are shared across markets and easily adaptable to every target market. This provides a generalized recommendation performance and a set of internal latent representations that are suitable for each individual market. Having such internal representations maximize the reusability of parameters translating into minimal effort on target market adaptation. To this end, we exploit the Model-Agnostic Meta-Learning (MAML) framework \cite{finn2017model} from the few-shot learning literature. \looseness=-1

\setlength{\textfloatsep}{0.2cm}
\begin{algorithm}[t]
\SetKwInput{KwInput}{Require}
\SetKwInput{KwOutput}{Require}
\DontPrintSemicolon

  \KwInput{Market set: $M=\{\mathbb{M}_0, \mathbb{M}_1,\cdots,\mathbb{M}_t\}$}
  \KwOutput{Step size parameters: $\alpha$, $\beta$; Number of shots: $K$; }
  
  Initialize NMF model parameters $\theta$ \;
  \While{not done}    
  { 
        \For{all $~\mathbb{M}_i \in M$}
        {
            adapt\_batch = Sample $K$ interactions from $\mathbb{M}_i$\\
            Evaluate: $\nabla_{\theta} \mathcal{L}_{\mathbb{M}_i}\text{NMF}(\theta)$ using $K$ adapt interactions\\
            Compute: $\theta'_{i} = \theta - \alpha \nabla_{\theta} \mathcal{L}_{\mathbb{M}_i}\text{NMF}(\theta)$\\
            eval\_batch = Sample another $K$ interactions from $\mathbb{M}_i$\\
        }
    $\theta = \theta - \beta \nabla_\theta \sum_{\mathbb{M}_i \in M} \mathcal{L}_{\mathbb{M}_i}\text{NMF}(\theta'_i)$ using eval batches\\
        
  }
  \For{all $~\mathbb{M}_i \in M$}
  {
     NMF$_i$: Fork a $\mathbb{M}_i$ specific NMF model\\
     Fine-tune NMF$_i$ using only the training data of $\mathbb{M}_i$ \\
  }
\caption{FOREC Training }
\label{alg:forec}
\end{algorithm}

The general neural architecture we use for our pre-training step is presented in middle part of Fig. \ref{fig:forec}. This architecture is first introduced by \citet{he2017neural} and widely used in the literature. Here, we summarize the Neural Matrix Factorization (NMF) deep network before explaining our learning paradigm across markets. NMF model fuses two sub-networks namely Generalized Matrix Factorization (GMF) and Multi-Layer Perceptron (MLP). Both GMF and MLP sub-networks are trained with the data individually and then fused using the NMF model architecture. For each user and item, a one-hot vector is constructed and fed to the user and item embedding layers of GMF and MLP networks, respectively. Note that GMF and MLP sub-networks keep their own embeddings space (i.e., no embedding sharing). Let the user and item latent vectors be $\mathbf{p}_u$ and $\mathbf{q}_i$, respectively. The GMF network simply perform an element-wise product between the vectors and uses a linear single-layer feed-forward network to calculate the prediction, i.e. $\mathbf{h}^{T} ( \mathbf{p}_u \odot \mathbf{q}_i)$. Here,  $\mathbf{h}$ denotes the weights of the output layer. For individual training of GMF, the prediction $\hat{y}_{\text{GMF}}$ is calculated using an activation function over the output layer's output. This implements a generalized matrix factorization network. 

For the MLP sub-network, the user and item vectors are concatenated and fed to a deep feed-forward network for learning the interactions between user and items, i.e. $z_1=\mathbf{p}_u \oplus \mathbf{q}_i$. Every layer of the deep network, takes the output of the previous layer $\mathbf{z}_k$ and calculates the output, i.e. $\mathbf{W}^T_{k}\mathbf{z}_{k-1} + \mathbf{b}_{k}$. $\mathbf{W}$ and $\mathbf{b}$ denote the weight matrix and bias vector of each layer. Here, we use ReLU as activation function and calculate $\hat{y}_{\text{MLP}}$ for individual training. The NMF model simply initialize the network with individual pre-trained parameters and for the output layer it concatenates the output layer of GMF and MLP with a hyper-parameter determining the trade-off. For the parameter initialization of our pre-training for market-agnostic NMF model, i.e. $\theta$ initialization, we simply concatenate the data from all target markets (a pair of markets in our study) and train the model. For the loss function, $\mathcal{L}$, we use the binary cross entropy between the target and the model output. 

Algorithm \ref{alg:forec} lines 2-10 present our market-agnostic NMF pre-training. The general framework for meta-learning considers a probability distribution over tasks. Given the highly imbalanced training data that each market offers and our final goal to learn a set of generalized network parameters that works for each individual market, we consider equal task sampling across markets. To this end, we simply iterate over each market instead of sampling randomly (line 3). During the meta-training with $K$-shot setting, for each market the model is trained such that it can adapt itself with only $K$ samples from the target market. This pushes the model parameters such that they broadly become applicable to each individual market. For this purpose, we sample two $K$ sized batches of the user-item interaction and perform adaptation and evaluation (lines 4 and 7).

Considering the NMF model with parameters $\theta$, with the adaptation step on market $\mathbb{M}_i$ the model parameters become $\theta'_i$. With a single gradient update, 
$\theta'_{i} = \theta - \alpha \nabla_{\theta} \mathcal{L}_{\mathbb{M}_i}\text{NMF}(\theta)$ in which $\alpha$ is the NMF model's learning rate (line 5-6). The meta-learning optimizes the NMF parameters across markets with the following meta-objective.
\begin{equation*}
    \min_{\theta} \sum_{\mathbb{M}_i \in M} \mathcal{L}_{\mathbb{M}_i}\text{NMF}(\theta'_{i})
\end{equation*}
The meta-optimization across markets are calculated using the $K$ evaluation user-item interactions on each market with a meta-learning step size, $\beta$. This updates the original model parameters such that few gradient steps can tune the parameters to a specific target market (line 9). 

\subsection{Forking: Market-Specific Model Construction}
After obtaining the general internal representations using our market-agnostic pre-training step, we conduct a series of operations for preparing a model that is market-specific for the desired target market. We call this step ``forking'' mainly due to the sharing of bottom parts and initialization of the middle parts of the market-specific model with the pre-trained model. The general schema of the forking operation is shown in Fig. \ref{fig:forec} for our FOREC algorithm. As it can be seen, from the market-agnostic NMF model trained on \market{us} and \market{de}, we fork market-specific NMF models for each of the models, shown on the right and left sides of the figure. 

Assuming the MLP part of the NMF model containing $m$ layers, and one additional layer for the NMF model, our objective with forking is to maximize the reusability of the general parameters. To this end, \citet{raghu2020rapid} studies the similarity between an adapted model's layers and the general model and suggests that the main body of the network barely changes and all the adaptation happens in the head layers of the network. Inspired by this finding, we freeze layers up to layer $k$ of the MLP network $(1 \leq k \leq m)$, the only layer of the GMF network, as well as the user and item embeddings learned with each of the sub-networks. Given that freezing some part of the network limits the capacity of the network for learning market-specific parameters, we add $n$ new market-specific layers on top of the original tower-style feed-forward network right after the NMF layer to increase the network's capacity. We call these new layers as MarketHead layers. We believe that our forking operation provides a network for balancing between the general market and target market-specific parameters after the final fine-tuning. The $k$ and $n$ values are experimentally explored on a few pairs and fixed for every market-specific forking in our experiments. The forking operation for each target market is shown in line $12$ of Algorithm \ref{alg:forec}. Further experimental details are given in Section \ref{sec:experimentalsetup}.

\subsection{Fine-Tuning: Final Training on the Target Market}
Over the forking step, we obtain a new market-specific NMF network for the desired target market that the bottom part of the network is frozen for any update for providing generalized internal features, middle part initialized with the general market that could easily adapted, and the final part that randomly initialized and needs further training. One advantage of such a design is that it facilitates the maintenance of the entire network with the availability of new interactions on other market(s). Having the new market-specific model forked and initialized as described, we simply fine-tune the model using only the data from the target market. We keep the loss function the same for this part. However, one could easily change the loss function with this step to better adapt the market needs in the target market. The fine-tuning operation for each target market is shown in line $13$ of Algorithm \ref{alg:forec}.

\section{Experiments}

\subsection{Experimental Setup} \label{sec:experimentalsetup}
\partitle{Dataset.} 
We use \xamazon dataset for our experimental evaluations. 
We specifically focus on \category{Elec.} category across $8$ markets presented in Table \ref{tab:stats}. We prepare our data similar to single-market experimental setup in the literature \cite{DBLP:conf/trec/AliannejadiMC16,DBLP:journals/tois/AliannejadiC18,zou2020towards,DBLP:journals/tkde/AliannejadiRC20,DBLP:conf/ecir/RahmaniABC20}. 
For the ratings, we filtered items and users that there exist less than five transactions.
We follow a long line of literature and use leave-one-out evaluation~\citep{cheng2016wide, he2017neural, kang2018self, hu2018conet, kang2019semi, ge2020learning, li2020ddtcdr}.

\newcommand{\bestsrc}{Best-Src\xspace}
\newcommand{\avesrc}{Ave-Src\xspace}
\newcommand{\fixsrc}{Fix-Src\xspace}
\newcommand{\staticrecomtablerowcount}{23}
\newcommand{\fullaugdamage}{13}
\newcommand{\selectiveaugdamage}{5}

\newcommand{\nsb}[1]{\textbf{#1}}       
\newcommand{\nos}[1]{{#1}}              
\newcommand{\msb}[1]{\textbf{#1\rlap{$^*$}}}   
\newcommand{\msn}[1]{{#1\rlap{$^*$}}}   
\newcommand{\psn}[1]{{#1\rlap{$^\dagger$}}}   
\newcommand{\psb}[1]{\textbf{#1\rlap{$^\dagger$}}}   
\newcommand{\fsn}[1]{{#1\rlap{$^\ddagger$}}}   
\newcommand{\fsb}[1]{\textbf{#1\rlap{$^\ddagger$}}}   

\setlength{\textfloatsep}{0.1cm}
\begin{table*}
    \centering
    \caption{Performance comparison of different \ac{CMR} methods. Best performing method in each source selection scenario is shown in bold fonts. Significance (Student's t-test) with $p<0.05$ compared to MAML and NMF++ is indicated by $^*$ and $^\dagger$, respectively.}
    \vspace{-0.3cm}
    \label{tab:performancecomparison}
    \resizebox{\textwidth}{!}{
        \begin{tabular}{lllllllllllllllllll}
        \toprule
        & & \multicolumn{8}{c}{\textbf{nDCG@10}} && \multicolumn{8}{c}{\textbf{HR@10}} \\
        \cmidrule(lr){3-9} \cmidrule(lr){11-19}
            & & \multicolumn{1}{c}{\market{de}} & \multicolumn{1}{c}{\market{jp}} & \multicolumn{1}{c}{\market{in}} & \multicolumn{1}{c}{\market{fr}} & \multicolumn{1}{c}{\market{ca}} & \multicolumn{1}{c}{\market{mx}} & \multicolumn{1}{c}{\market{uk}} &  & \multicolumn{1}{c}{\market{de}} & \multicolumn{1}{c}{\market{jp}} & \multicolumn{1}{c}{\market{in}} & \multicolumn{1}{c}{\market{fr}} & \multicolumn{1}{c}{\market{ca}} & \multicolumn{1}{c}{\market{mx}} & \multicolumn{1}{c}{\market{uk}} \\
        \cmidrule(lr){1-19}
\multirow{3}{*}{\rotatebox[origin=c]{90}{Single}}   
&GMF        & \nos{0.1109} & \nos{0.2016} & \nos{0.0324} & \nos{0.4085} & \nos{0.2960} & \nos{0.5626} & \nos{0.4685} && \nos{0.4435} & \nos{0.3033} & \nos{0.0843} & \nos{0.5246} & \nos{0.4500} & \nos{0.6452} & \nos{0.5923} \\
&MLP        & \nos{0.2506} & \nos{0.3331} & \nos{0.5921} & \nos{0.3980} & \nos{0.2522} & \nos{0.5368} & \nos{0.4662} && \nos{0.4642} & \nos{0.4324} & \nos{0.6867} & \nos{0.5424} & \nos{0.4444} & \nos{0.6636} & \nos{0.5978} \\
&NMF        & \nos{0.2486} & \nos{0.3394} & \nos{0.5970} & \nos{0.3961} & \nos{0.2917} & \nos{0.5212} & \nos{0.4853} && \nos{0.4824} & \nos{0.4447} & \nos{0.6807} & \nos{0.5451} & \nos{0.4533} & \nos{0.6648} & \nos{0.6035} \\
\cmidrule(lr){1-19}\morecmidrules\cmidrule(lr){1-19}
\multirow{8}{*}{\rotatebox[origin=c]{90}{\bestsrc}} 
&GMF++      & \nos{0.2809} & \nos{0.3253} & \nos{0.5798} & \nos{0.4074} & \nos{0.2853} & \nos{0.5598} & \nos{0.4628} && \nos{0.4605} & \nos{0.4713} & \nos{0.6807} & \nos{0.5532} & \nos{0.4411} & \nos{0.6581} & \nos{0.5945} \\
&MLP++      & \nos{0.2643} & \nos{0.3268} & \nos{0.5891} & \nos{0.3991} & \nos{0.2715} & \nos{0.5557} & \nos{0.4801} && \nos{0.4753} & \nos{0.4939} & \nos{0.6807} & \nos{0.5604} & \nos{0.4389} & \nos{0.6642} & \nos{0.6053} \\
&NMF++      & \nos{0.3116} & \nos{0.3890} & \nos{0.6020} & \nos{0.4165} & \nos{0.3134} & \nos{0.5511} & \nos{0.5050} && \nos{0.5083} & \nos{0.5123} & \nos{0.6988} & \nos{0.5742} & \nos{0.4692} & \nos{0.6765} & \nos{0.6232} \\
&DDTCDR         & \nos{0.2553} & \nos{0.2960} & \nos{0.4994} & \nos{0.3702} & \nos{0.2958} & \nos{0.5182} & \nos{0.4192} && \nos{0.4245} & \nos{0.3873} & \nos{0.5542} & \nos{0.5228} & \nos{0.4717} & \nos{0.6028} & \nos{0.5855} \\
&MAML       & \psn{0.3402} & \psb{0.4258} & \nos{0.6076} & \psn{0.4686} & \psn{0.3512} & \psn{0.5868} & \psb{0.5272} && \psn{0.5306} & \psn{0.5635} & \nos{0.6988} & \psn{0.6046} & \psn{0.5058} & \psn{0.7066} & \psb{0.6478} \\
 
\cmidrule(lr){2-19} 
 
&NMF-\ourmodel & \psn{0.3515} & \nos{0.3847} & \psn{0.6249} & \psn{0.4387} & \psn{0.3454} & \psn{0.5886} & \nos{0.5102} && \nos{0.5169} & \nos{0.5082} & \nos{0.7048} & \nos{0.5748} & \nos{0.4735} & \nos{0.6832} & \nos{0.6150} \\ 
&\ourmodel  & \msb{0.3621} & \psn{0.4195} & \msb{0.6378} & \psb{0.4755} & \psb{0.3693} & \msb{0.6160} & \psn{0.5252} && \psb{0.5480} & \psb{0.5717} & \msb{0.7169} & \psb{0.6148} & \psb{0.5159} & \psb{0.7152} & \psn{0.6465} \\

\cmidrule(lr){1-19}\morecmidrules\cmidrule(lr){1-19}

\multirow{8}{*}{\rotatebox[origin=c]{90}{\avesrc}} 
&GMF++          & \nos{0.2707} & \nos{0.3036} & \nos{0.4986} & \nos{0.4005} & \nos{0.2827} & \nos{0.5360} & \nos{0.4587} && \nos{0.4537} & \nos{0.4198} & \nos{0.5813} & \nos{0.5412} & \nos{0.4385} & \nos{0.6442} & \nos{0.5882} \\
&MLP++          & \nos{0.2573} & \nos{0.3350} & \nos{0.5667} & \nos{0.3970} & \nos{0.2619} & \nos{0.5352} & \nos{0.4688} && \nos{0.4657} & \nos{0.4588} & \nos{0.6649} & \nos{0.5495} & \nos{0.4339} & \nos{0.6552} & \nos{0.5963} \\
&NMF++          & \nos{0.2876} & \nos{0.3533} & \nos{0.5809} & \nos{0.4123} & \nos{0.3044} & \nos{0.5387} & \nos{0.4899} && \nos{0.4890} & \nos{0.4685} & \nos{0.6687} & \nos{0.5595} & \nos{0.4608} & \nos{0.6652} & \nos{0.6096} \\
&DDTCDR         & \nos{0.2155} & \nos{0.2044} & \nos{0.3095} & \nos{0.3285} & \nos{0.2551} & \nos{0.4670} & \nos{0.3923} && \nos{0.3678} & \nos{0.2945} & \nos{0.4053} & \nos{0.4560} & \nos{0.4117} & \nos{0.5675} & \nos{0.5416} \\
&MAML           & \psn{0.3336} & \nos{0.3756} & \nos{0.5948} & \psn{0.4555} & \psn{0.3463} & \psn{0.5793} & \psn{0.5174} && \psn{0.5111} & \nos{0.4941} & \nos{0.6764} & \psn{0.5959} & \psn{0.4939} & \psn{0.6878} & \psn{0.6335} \\

\cmidrule(lr){2-19}

&NMF-\ourmodel  & \psn{0.3340} & \nos{0.3742} & \nos{0.5974} & \psn{0.4374} & \psn{0.3423} & \psn{0.5747} & \nos{0.5063} && \nos{0.4968} & \nos{0.4819} & \nos{0.6713} & \nos{0.5652} & \nos{0.4673} & \nos{0.6657} & \nos{0.6079} \\  
&\ourmodel      & \msb{0.3523} & \msb{0.4007} & \psb{0.6140} & \msb{0.4637} & \msb{0.3616} & \msb{0.5989} & \psb{0.5200} && \msb{0.5280} & \msb{0.5187} & \psb{0.6937} & \psb{0.5994} & \msb{0.5064} & \psb{0.6981} & \psb{0.6364} \\

\cmidrule(lr){1-19}\morecmidrules\cmidrule(lr){1-19}

\multirow{8}{*}{\rotatebox[origin=c]{90}{\fixsrc (\market{us})}} 
&GMF++          & \nos{0.2571} & \nos{0.3123} & \nos{0.5649} & \nos{0.3844} & \nos{0.2825} & \nos{0.5162} & \nos{0.4628} && \nos{0.4498} & \nos{0.4201} & \nos{0.6687} & \nos{0.5382} & \nos{0.4398} & \nos{0.6311} & \nos{0.5945} \\
&MLP++          & \nos{0.2566} & \nos{0.3227} & \nos{0.5728} & \nos{0.3935} & \nos{0.2773} & \nos{0.5291} & \nos{0.4693} && \nos{0.4498} & \nos{0.4344} & \nos{0.6747} & \nos{0.5547} & \nos{0.4347} & \nos{0.6519} & \nos{0.6026} \\
&NMF++          & \nos{0.3008} & \msn{0.3446} & \nos{0.6020} & \nos{0.4208} & \nos{0.3101} & \nos{0.5509} & \nos{0.4994} && \nos{0.4857} & \nos{0.4508} & \nos{0.6988} & \nos{0.5598} & \nos{0.4610} & \nos{0.6630} & \nos{0.6151} \\
&DDTCDR         & \nos{0.2376} & \nos{0.2196} & \nos{0.3763} & \nos{0.3702} & \nos{0.2958} & \nos{0.3592} & \nos{0.4192} && \nos{0.3997} & \nos{0.3299} & \nos{0.4277} & \nos{0.5228} & \nos{0.4717} & \nos{0.5433} & \nos{0.5855} \\
&MAML           & \psn{0.3295} & \nos{0.3154} & \nos{0.5622} & \nos{0.4403} & \psn{0.3512} & \psn{0.5970} & \nos{0.5140} && \nos{0.5065} & \nos{0.4488} & \nos{0.6506} & \psn{0.5844} & \psn{0.5058} & \psn{0.7035} & \psn{0.6315} \\

\cmidrule(lr){2-19}

&NMF-\ourmodel  & \psn{0.3265} & \msb{0.3620} & \msb{0.6249} & \psn{0.4417} & \psn{0.3394} & \nos{0.5671} & \nos{0.5076} && \nos{0.4879} & \nos{0.4549} & \nsb{0.7048} & \nos{0.5658} & \nos{0.4635} & \nos{0.6593} & \nos{0.6109} \\
&\ourmodel      & \psb{0.3306} & \msn{0.3563} & \msn{0.6143} & \psb{0.4485} & \msb{0.3693} & \psb{0.6160} & \psb{0.5252} && \psb{0.5158} & \msb{0.4877} & \nos{0.6928} & \psb{0.5886} & \psb{0.5159} & \psb{0.7152} & \psb{0.6420} \\

\bottomrule
        \end{tabular}
    }
\end{table*}

\partitle{Compared methods.}
\label{sec:experimentalruns}
In order to show the effectiveness of our method, we employ the following models on each target market:
\begin{itemize}[leftmargin=*]
    \item {\bfseries GMF, MLP, NMF:} The \ac{GMF}, \ac{MLP} and \ac{NMF} models from~\citep{he2017neural} trained using only the target market. 
    \item {\bfseries GMF++, MLP++, NMF++:} The simplest way of leveraging the cross-market data is to train the model on the interactions inside both the source and target markets by sharing the item and user representations. 
    We equally sample from both markets in the training phase---equal number of training interactions from each market is used. We experimentally observed that this training provides higher performance compared to simple concatenation of both markets. \looseness=-1
    \item {\bfseries DDTCDR:} \ac{CDR} and \ac{CMR} have similarities as discussed in Sec.~\ref{sec:relatedwork}. To test whether high performing \ac{CDR} methods can be used to effectively solve the \ac{CMR} problem, we include the algorithm proposed in~\citep{li2020ddtcdr}, as one of the recent strong \ac{CDR} methods in our comparison. As the assumption with \ac{CDR} is that the set of users are shared across two domains, the original model connects the user features between a pair of MLP networks using an orthogonal transformation matrix. We adapt the model into \ac{CMR} by connecting the item features between two market's MLP networks. We performed a similar modifications for the CoNet's network structure proposed by \citet{hu2018conet} for the \ac{CDR} problem. We only report DDTCDR due to its consistent superiority compared to CoNet and the space limitations. 
    
    \item {\bfseries MAML:} Meta-learning is widely used in the recommendation literature for variety of problem settings---see Section \ref{sec:relatedwork}. Here, we adapt the learning paradigm to the \ac{CMR} by employing the \ac{MAML} framework providing model-agnostic solution for meta-learning. Our \ac{MAML} training phase is described in Algorithm \ref{alg:forec} lines 1-10. Here, after the training phase, we perform a single pass with $K$-shots of sampled interactions from the validation split of the target market and fast adapt the pre-trained model parameters to the target market. We observed no further improvements with more number of passes on the adaptations. This baseline provides the sole importance of our \ac{MAML} adaptation to the \ac{CMR} problem. 
    
    \item {\bfseries NMF-FOREC:} In order to show the impact of our MAML-based pre-training with the FOREC, we pre-train our market-agnostic model with only NMF++ method described above and perform forking and fine-tuning for each specific market. This baseline provides evidences in two ways; (1) The importance of MAML pre-training on the performance of the FOREC model, (2) The sole impact of forking and fine-tuning operations over a weak pre-training of internal features.
\end{itemize}

\partitle{Hyper-parameters.}
For \ac{GMF}, \ac{MLP} and \ac{NMF} we follow~\citep{he2017neural} and set all the network structure and the latent factor dimension as suggested, i.e. $[16,64,32,16,8]$ with $8$ as embedding dimensions. 
For the optimizer, we use Adam~\citep{kingma2014adam} and select the learning rate and $l_2$-regularization coefficient hyper-parameters using the validation data of a subset of our markets. For the learning rate we considered $\{0.1, 0.05, 0.01, 0.005, 0.001, 0.0001\}$ and selected $0.01$ for MLP and NMF and $0.005$ for the GMF model. For regularization we observed that $1e^{-7}$ is the best among our consideration set. We use $4$ negative training samples for each user re-sampled with each epoch. ReLU is used as the activation function. 
For DDTCDR we use the same hyper-parameters provided in the original implementation of~\citep{li2020ddtcdr}. As the model uses a preset embeddings, we employ the GMF model for the initialization. For MAML training, we set the fast learning rate $\beta=0.1$ selected from $\{0.5, 0.3, 0.1, 0.01, 0.001\}$ and the number of shots as $20$ selected from $\{5, 20, 50, 100, 200\}$. 
For our FOREC's market-agnostic part we use the same architecture as of NMF and employ the last three layers of the NMF network for forking and freeze the remaining bottom layers as well as the embeddings. For the market head layers, we considered three different layer sizes; (a) no new layers, (b) two $16$ layers, and (c) 3 layers with $[16, 32, 16]$, and evaluated experimentally using the validation data on a subset of markets. We selected (c) as our model's market head layers for all experiments. In addition, we observed that setting higher $l_2$-regularization with the fine-tuning step helps the overall performance, especially with lower resourced markets---we set it to $0.001$ for all fine-tuning steps.

\partitle{Evaluation Metrics.} 
We use \ac{HR} and nDCG as our evaluation metrics, commonly used in the literature. We report these metrics for a cut-off of 10. Similar to other works, we constructed the ground truth using the buying behavior by considering an item as relevant if the user gave a rating. In addition, we follow the literature and sample 99 negative items for each user in our evaluations. \looseness=-1

\vspace{-0.5em}
\subsection{Results and Discussion}
We compare our~\ourmodel model to several baseline techniques discussed in Sec.~\ref{sec:experimentalsetup} in terms of recommendation performance.
In theory, each target market can be paired with each auxiliary source market.
However, for the $7$ markets that we consider, all possible pairings of source and target markets leads to $49$ different settings.
For better readability and due to space limitation, we report our results in the following scenarios for every target market:
\begin{itemize}[leftmargin=*]
    \item {\bfseries \bestsrc:} Each of the parallel markets are once considered as source and the model is evaluated using the nDCG@10 on the validation set. For each method, the source market with the highest improvement on the validation set is selected. As such, for a single target market, the best source market for different \ac{CMR} methods may be different.
    \item {\bfseries \avesrc:} In order to provide a rough indicator of the safer choice of the \ac{CMR} method on each target market along with an overall insight on different source selections, we report the average performance of each model using different source markets.
    \item {\bfseries \fixsrc:} We report the results when a fixed base market is available (i.e. \market{us} market) and all \ac{CMR} methods can only use that market to improve the target market recommendation.
\end{itemize}
Table \ref{tab:performancecomparison} summarizes the evaluation results with and without cross-market data. All three aforementioned source selection scenarios are reported respectively in the Table. The best performing method for each target market and each source selection scenario is shown by bold fonts.

\setlength{\textfloatsep}{0.2cm}
\begin{figure}[t]
    \centering
    \includegraphics[width=\columnwidth]{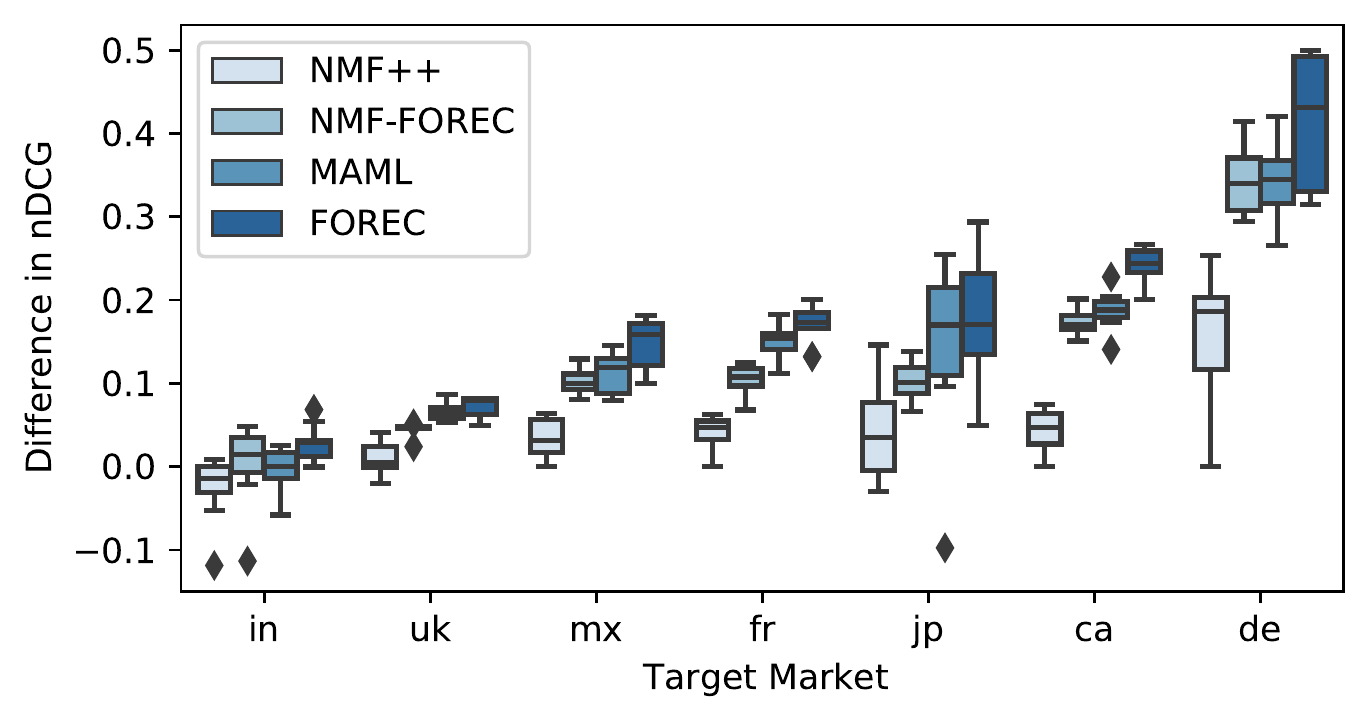}
    \vspace{-2.5em}
    \caption{Impact of choosing different source markets on different target markets for different models.}
    \label{fig:delta_ndcg}
\end{figure}

We observe that~\ourmodel is the winner in almost all target markets and across all the source selection scenarios, except for \market{jp} and \market{uk} in \bestsrc scenario, and \market{jp} and \market{in} for \fixsrc. In addition, we observe that even our simple baselines are able to utilize to some degree the cross-market data provided and improve the target market's recommendation performance. This suggests the importance of cross-market training, even with a fixed base market, for better recommendation systems across different local markets. It also suggests the importance of the model and source market selection for deployment purposes.   

When the source market is selected by the validation performance (i.e. \bestsrc scenario), \ourmodel and \ac{MAML} noticeably outperform other \ac{CMR} methods. 
For \market{uk}, \ac{MAML} is slightly better than \ourmodel, though the difference is not substantial. For \market{jp}, \ac{MAML} has a higher nDCG, but \ourmodel is better in terms of \ac{HR}, suggesting that neither can be picked as the winner. For the other $5$ markets, \ourmodel outperforms \ac{MAML} both in terms of nDCG and \ac{HR}.\looseness=-1

Looking at the results in \avesrc scenario, \ourmodel outperforms other powerful \ac{CMR} methods on average, meaning that given a source and target market, \ourmodel is a safer choice for \ac{CMR}.
This is further illustrated in Table~\ref{tab:winnerloser} where we report for each target market the number of source markets for which each of \ourmodel or \ac{MAML} is the winner along with the relative percentage improvement of \ourmodel compared to \ac{MAML}.
As it can be seen, over all $49$ possible combinations of source-target pairs, only in $5$ of them \ac{MAML} is slightly better than \ourmodel.
As source markets, \market{mx} and \market{uk} are the ones with the highest average improvement of \ourmodel over MAML, while \market{de} market is the one where MAML performs closest to \ourmodel and even surpasses it in two target markets.

Comparing the \fixsrc results, suggests that even though \market{us} is not the best source market for all target markets in terms of performance boost (see \bestsrc scenario), comparing with the single market results, it can be observed that \market{us} helps all of the tested target markets and the \ac{CMR} comes to its highest performance when used with our \ourmodel model. In addition, we notice that NMF-FOREC is performing better in two markets compared to \ourmodel. We hypothesize that this might be due to the relative data size differences between the \market{us} and each of these markets (see Table \ref{tab:marketstats}), performing the pre-training using MAML provides a limited added value. \looseness=-1 

\setlength{\textfloatsep}{0.2cm}
\begin{table}[t]
    \caption{nDCG Comparison of \ourmodel vs. \ac{MAML} with different source markets. Values denote the percentage of relative improvement of \ourmodel compared with \ac{MAML}. Positive values indicate \ourmodel superiority.}
    \vspace{-0.3cm}
    \label{tab:winnerloser}
    \resizebox{\columnwidth}{!}{
    \begin{tabular}{ccrrrrrrr|c}
    \toprule
    && {\market{de}} & {\market{jp}} & {\market{in}} & {\market{fr}} &  {\market{ca}} &  {\market{mx}} &  {\market{uk}} & {\market{avg}}\\
    \cmidrule{2-10}
    \multirow{8}{*}{\rotatebox[origin=c]{90}{{Source Markets}}}
    & \market{de}             &     -- &  -1.62 &  4.96 & 4.07 & 3.55 & 1.23 & -0.74 & 1.64\\
    & \market{jp}             &  7.33 &      -- &  2.87 & 1.05 & 0.40 & 3.39 &  0.06 & 2.16\\
    & \market{in}             &  2.56 &   6.48 &     -- & 0.75 & 4.90 & 2.94 & -0.40 & 2.46\\
    & \market{fr}             & 14.34 &  -1.47 &  1.42 &    -- & 7.67 & 1.98 &  0.05 & 3.43\\
    & \market{ca}             &  9.81 &   5.75 & -0.08 & 0.32 &    -- & 3.44 &  0.62 & 2.84\\
    & \market{mx}             &  4.51 &   6.24 &  3.48 & 2.86 & 5.28 &    -- &  0.79 & 6.18\\
    & \market{uk}             &  3.87 &  24.51 &  2.39 & 3.99 & 4.86 & 3.62 & --    & 6.18\\
    & \market{us}             &  0.32 & 12.97 &  9.26 & 1.85 & 5.16 & 3.18 &  2.17 & 4.99\\
        \cmidrule{2-10}
    & \market{avg}            & 5.34 & 6.61 & 3.04 & 1.86 & 3.98	 & 2.47 & 0.32    & --\\
    \bottomrule
    \end{tabular}
    }
\end{table}

Comparing NMF-\ourmodel with \ourmodel, we observe that in most of the target markets, \ourmodel performs better. Based on the observations from Table~\ref{tab:performancecomparison}, in most of the cases, MAML performs better than NMF++ and \ourmodel performs better than MAML. This shows the significance of the MAML-based pre-training as well as forking and fine-tuning, as proposed by our model.

Fig.~\ref{fig:delta_ndcg} compares these four models more deeply.
In this figure, the nDCG@10 improvement of these \ac{CMR} models over the NMF on single market are shown for different target markets.
For each method in each target market, the distribution of nDCG@10 improvements based on different source markets is given as a box plot.
This figure provides a better illustration of the trend between these four models described earlier.

Finally, we observe that the adopted DDTCDR, as one of the state-of-the-art \ac{CDR} methods, does not perform well for \ac{CMR}.
In order to adopt a \ac{CDR} method to \ac{CMR} scenario, as described before, the users and items should be interchanged: the shared users in \ac{CDR} are analogous to the shared items in \ac{CMR}.
This change of perspective looks natural at first sight, but it introduces some issues.
Here we discuss the issue.
In the item recommendation problem, a number of past interactions of a user with items are used to predict her future interactions.
This means that the evaluation is based on the accuracy of the predicted items for users.
In \ac{CDR}, the users are shared across domains and the interacted items from the source domain add to the per-user information of the target domain.
Differently, in \ac{CMR}, the items are shared across markets and the users are separate.
Changing the perspective of users/items during training but having a fixed evaluation based on the per-user predictions is the issue of na\"ively adopting a \ac{CDR} to \ac{CMR}.
An interesting future direction would be to analyze the impact of such adoption for user recommendation in \ac{CMR} scenarios.

\vspace{-0.5em}
\subsection{Impact of Cross-Market Training on Different Users}
\begin{figure}
    \centering
    \vspace{-5mm}
    \subfloat[User group.]{
    \includegraphics[width=0.5\columnwidth]{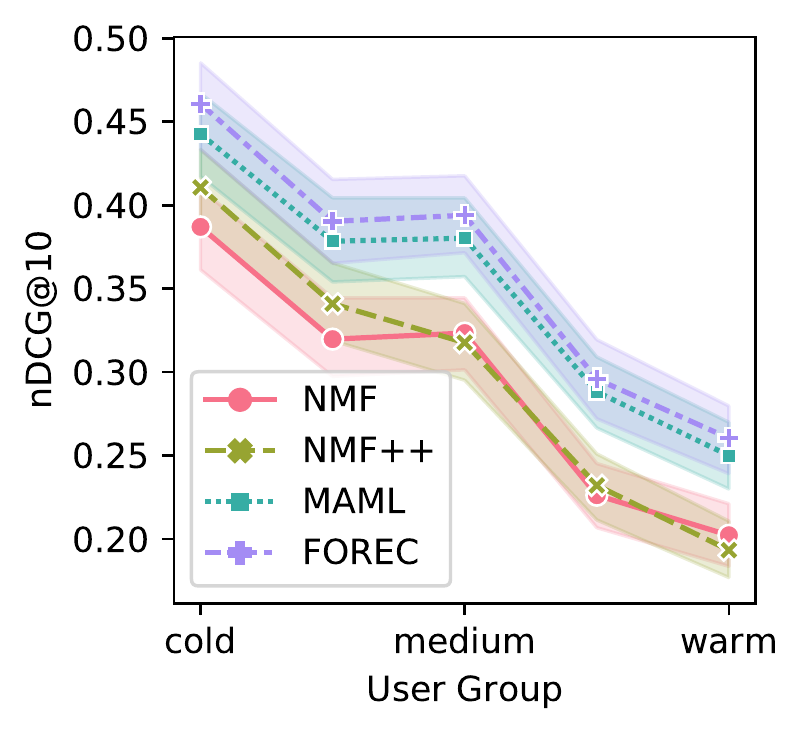}\label{fig:per_user}
    }
    \subfloat[Data size.]{
    \includegraphics[width=0.487\columnwidth]{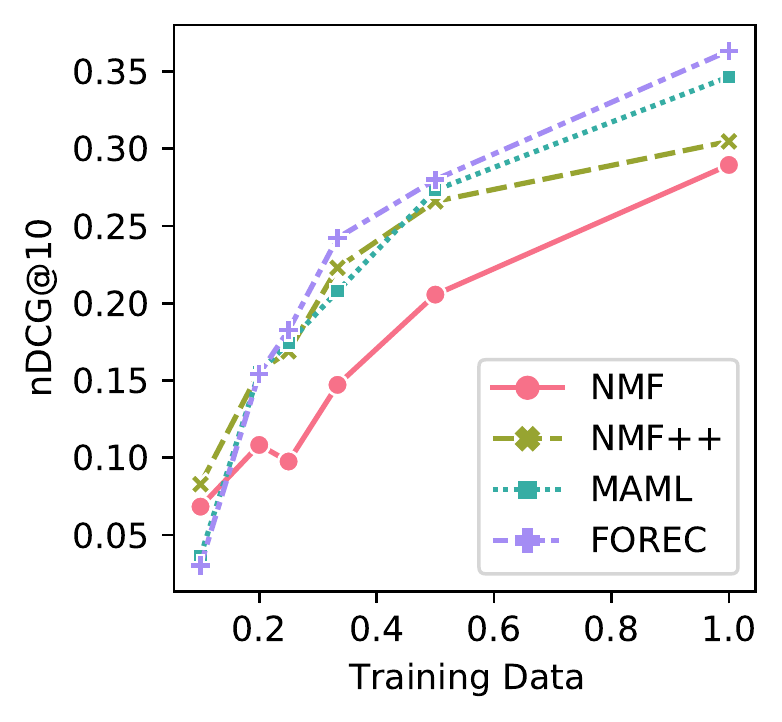}\label{fig:target_size}
    }
    \vspace{-2.5mm}
    \caption{Performance comparison of FOREC on (a) different user groups; and (b) different training data size.
        }
        \vspace{-3mm}
\end{figure}

Here we study the impact of different cross-market training approaches on user groups in terms of their training data size.
Following the work of \citet{DBLP:journals/pvldb/LiuPCY17} we split the users based on the interactions into five groups. The users with the least number of interactions are named \textit{cold} (average of 5 interactions), and the ones with the highest number of interactions are called \textit{warm} (average of 13.3 interactions). We create five equally-sized user splits and report the average nDCG@10 for each group in Fig.~\ref{fig:per_user}.
This figure contains the performance of four models on these five user groups in the \market{ca} market: single market NMF, NMF++, MAML and FOREC.
Here, we used \market{de} as the source market.\footnote{We repeated the experiments with different source and target markets and got similar results. Here we only show one case.}
We observe that MAML and \ourmodel almost uniformly improve the performance over the single market NMF in all five user groups.
Comparing \ourmodel with MAML, again we see a consistent improvement in all five user groups with a slightly bigger gap toward the cold user group.
This observation, together with other similar observations on our tests over other source-target pairs, provide experimental evidence that \ourmodel is suitable both for cold- and warm-start situations in the target market.
NMF++, on the other hand, only helps the cold user groups, as can be seen in the figure.

\setlength{\textfloatsep}{0.4cm}

\vspace{-0.5em}
\subsection{Impact of the Target Market Data Size}
In order to study the impact of target training data size, for a given market pair, we gradually decrease the number of training interactions for each user. 
Starting from the full target market user-item interactions, each time we halve the training data for the target market until only $10\%$ of the target market's training interactions remains for each user. 
We train four of our models (NMF, NMF++, MAML, \ourmodel) on each of these settings and test on the target market. Fig. \ref{fig:target_size} presents the resulting nDCG@10 performance on \market{ca} with \market{uk} as the source market---similar observations made for a many pairs. 
As it can be seen, among the cross-market methods we see that \ourmodel and MAML are performing similarly especially when the target market's size is extremely small. On the extremely small target market size ($10\%$ of the data), we see that single market NMF model as well as NMF++ are performing better. As more data becomes available, a significant boost among cross-market methods are observed, especially with \ourmodel---as opposed to NMF with lazy reaction to the new data availability. With further data availability, we see the full superiority of MAML as well as \ourmodel. This analysis suggests that some minimum amount of training data in the target market is essential for the cross-market models that we examined in order to be able to make use of the auxiliary source market. We hypothesize that when the target market provides limited amount of training data the pre-training through the MAML approach shifts parameters more toward the source market compared to the NMF++ training approach. However, the general observation from this analysis could be that \ourmodel and MAML are relatively resilient to the amount of the target training data. We note that this requires further analysis which is out of the scope of our study.\looseness=-1 

\section{Conclusion \& Future Work}
We studied the problem of recommending relevant products to users in relatively resource-scarce markets by leveraging data from similar or richer-in-resource auxiliary markets. To this aim, we introduced a large-scale real-life dataset, named as XMarket, providing product information and reviews on $18$ Amazon marketplaces featuring $52.5$ million user-item interactions. We hypothesized and showed through extensive experiments on $7$ target markets that data from one market can actually be used to improve the performance in another. Our model, named as FOREC, demonstrates robust effectiveness, consistently improving the performance on target markets compared to competitive baselines selected for our analysis. 
In particular, FOREC improves on average 24\% and up to 50\% in terms of nDCG@10, compared to the NMF baseline.    

Our analysis and experiments suggest specific future directions in this research area. We show that models that are designed for \ac{CDR} are not necessarily suitable for the market adaptation problem setting. One interesting extension of our study could be designing models that are domain and market agnostic in the sense that they can consume the data across different markets and domains and leverage that for the improved recommendation on a target market's specific domain. In addition, we believe that data filtering or selection across markets could potentially be helpful for the models we discussed in our study. Moreover, using data augmentation techniques to generate synthetic ratings for target markets \cite{chae2020ar,wang2019enhancing} could be a potential solution for the extreme low-resource markets, i.e. cold-start markets. We believe that many potentially interesting problems are yet to be explored in the \ac{CMR} area. 

\partitle{Acknowledgments.}
    This work was supported in part by the Center for Intelligent Information Retrieval and in part by 
    the NWO Innovational Research Incentives Scheme Vidi (016.Vidi.189.039),
    the NWO Smart Culture - Big Data / Digital Humanities (314-99-301), 
    the Elsevier and NWO (612.\-001.\-551), and
    the H2020-EU.3.4. - SOCIETAL CHALLENGES - Smart, Green And Integrated Transport (814961). Any opinions, findings and conclusions or recommendations expressed in this material are those of the authors and do not necessarily reflect those of the sponsors. 

\clearpage
\bibliographystyle{ACM-Reference-Format}
\balance
\bibliography{myref}

\end{document}